# Classical Option Pricing and Some Steps Further

Victor Olkhov

TVEL, Moscow, Russia

victor.olkhov@gmail.com

ORCID: 0000-0003-0944-5113

## ABSTRACT

This paper considers the asset price $p$ as relations $C=pV$ between the value $C$ and the volume $V$ of the executed transactions and studies the consequences of this definition for the option pricing equations. We show that the classical BSM model implicitly assumes that value $C$ and volume $V$ of transactions follow identical Brownian processes. Violation of this identity leads to 2-dimensional BSM-like equation with two constant volatilities. We show that agents expectations those approve execution of transactions can further increase the dimension of the BSM model. We study the case when agents expectations may depend on the option price data and show that such assumption can lead to the nonlinear BSM-like equations. We reconsider the Heston stochastic volatility model for the price determined by the value and the volume and derive 3-dimensional BSM-like model with stochastic value volatility and constant volume volatility. Variety of the BSM-like equations states the problem of reasonable balance between the accuracy and the complexity of the option pricing equations.

Keywords: Option Pricing, Black-Scholes-Merton Equations, Stochastic Volatility, Market Transactions, Expectations, nonlinear equations

JEL: C10, G110, G120, G130

# 1. Introduction.

We consider classical option pricing (Black and Scholes 1973; Merton 1973) Black-Schole-Merton (BSM) equation, non-linear and stochastic volatility (Heston 1993) models and study some extensions. Classical papers by Black, Scholes (1973) and Merton (1973) were published almost 50 years ago but nevertheless their results support current development and further research of assets and option pricing models. A vast amount of researches contribute to this important financial problem. We refer only few papers by Bates (1996), Merton (1997), Scholes (1997) and some who contribute to general treatment of options pricing and financial methods - Figlewski (1998), Shiryaev (1999), Hull (2009). Many researchers provide extension of the BSM model: Cox and Huang, (1987), Hull and White (2001). Extensions of BSM from constant to stochastic volatility were developed by Hull and White (1987), Heston (1993), Ball and Roma (1994), Saikat (1996), Britten-Jones and Neuberger (2000), Engle and Figlewski (2014), Cohen and Tegner (2018), multiple assets option pricing models by Broadie and Detemple (1997), Rapuch and Roncalli (2004), Carmona and Durrleman (2006), Li, Deng and Zhou, (2010), application of Non-Gaussian processes by Borland (2004), extension of diffusion by Kleinert and Korbel (2016). Some collections of different approaches to BSM model are presented in Choi (2018). We have no intend to give any reasonable review of current state of derivatives and option pricing theory. Instead we indicate some extensions of classical BSM model – extension from constant to stochastic volatility models, multiple assets options pricing, non-Brownian random processes and etc. It seems that during these 50 years after Black and Scholes (1973), and Merton (1973) studies almost all possible methods and directions for options pricing modeling are already described.

Nevertheless we regard the classical BSM model as an infinite source for further development. Time by time it is useful to reconsider and reevaluate the assumptions underlying the classical models. It may help to find out the way for further progress. In this paper we consider the classical BSM model and state a simple question – what economic factors define price evolution and how that match the assumption on Brownian behavior of the underlying asset's price?

Indeed, the asset price is not the March Hare from Lewis Carroll's "Alice's Adventures in Wonderland" that can jump randomly. The price is not a stand-alone financial notion and don't behave arbitrary like "The Cat that walked by himself" by R. Kipling. The price dynamics is determined by numerous economic and financial factors. Stochastic behavior of these factors impact stochastic behavior of the price. Option pricing models should follow the general relations between economic and financial variables and market transactions that determine random market evolution and define stochastic price dynamics. Otherwise some studies on option pricing seem alike to math gambling to guess the correct form for the stochastic process that govern the price or its volatility without efforts to understand financial reasons of such dynamics. Numerous such attempts improve classical BSM model and enhance the option price studies but give a little for the understanding – what are the financial relations that govern the assets price randomness?

This paper use common price definition as result of executed market transactions. Researchers use numerous definitions of price. More than a century ago Fetter (1912) mentioned 117 price definition and one of them define the price as "Ratio-of-exchange definitions of price in terms of value in the sense of a mere ratio of exchange". In this paper we use this particular price definition to describe price random properties and option pricing. This price notion is well known and defines the price $p$ of the transaction as coefficient between the total value $C$ and the total volume $V$ of the transaction:

$$C = pV \qquad (1.1)$$

It is obvious that other numerous price definitions due to Fetter (1912) and their usage for modeling or divinations of asset price dynamics, as endogenous or exogenous price process will remain in use for a long time. We underline that any such price fortune-telling have sense as ground for price expectations only. Only definition (1.1) establishes the single source of the confirmed data series for the price of executed transactions. Obviously various price expectations in agents minds impact agents decisions to go into particular transaction (1.1). Economic agents with numerous price expectations, value expectations, volume expectations and so on execute market transactions. The trading data of the transactions executed under agents expectations deliver the (1.1) as confirmed price dynamics. We take relations (1.1) as the only ground for option pricing. However, it is clear that agents expectations that approve transactions impact the price properties. Below we show how simple relations (1.1) between the value $C$, the volume $V$ and the price $p$ of the market transactions and considerations of expectations as factors that impact trading records allow contribute to the BSM, non-linear and stochastic volatility option pricing models.

In Sec.2 we discuss the BSM model taking into account (1.1.). In Sec.3 we show that impact of agents expectations can give further complexities of the BSM model. In Sec.4 we discuss non-linear option pricing as result of impact of expectations. In Sec. 5 we discuss Heston's stochastic volatility model taking into account relations (1.1). In Sec. 6 and 7 we present Discussion and Conclusion. Equation (3.2) defines Section 3 and equation 2.

## 2. Market transactions and option pricing

Let's take the classical BSM model (Black and Scholes 1973; Merton 1973; Hull 2009) and assume that the underling asset price $p(t)$ follows the Brownian motion $dW(t)$ as:

$$dp(t) = p(t)[\mu\, dt + \sigma dW(t)] \tag{2.1}$$

$$<dW(t)> = 0; \quad <dW(t)dW(t)> = dt \tag{2.2}$$

Here $\mu$ – linear trend, $\sigma$ - dispersion and $r$ – risk-free rate and $\mu, \sigma, r$ – are constant. We use notion <...> to define averaging of random process. Option price $S(t,p)$ follows the classical BSM equation (2.3):

$$\frac{\partial S}{\partial t} + rp\frac{\partial S}{\partial p} + \frac{1}{2}\sigma^2 p^2 \frac{\partial^2 S}{\partial p^2} = rS \tag{2.3}$$

Let's maintain all assumptions of the classical BSM model except the main one: we don't take assumptions (2.1, 2.2) on the price $p(t)$. We suppose that stochastic behavior of the price $p(t)$ should be determined by random properties of market transactions with underling assets and in particular by random dynamics of the value $C(t)$ and the volume $V(t)$ of the transactions. Trivial relations define the value $C(t)$ and the volume $V(t)$ of the market transaction $M(t)$ and the price $p(t)$ as:

$$M(t) = \bigl(C(t), V(t)\bigr); \quad p(t) = \frac{C(t)}{V(t)} \tag{2.4}$$

Relations (2.4) are trivial but they replace initial assumptions (2.1, 2.2) on random properties of the price $p(t)$ by assumptions on random properties of the value $C(t)$ and the volume $V(t)$ of the transaction $M(t)$. To keep simplicity of BSM model as a first approximation let's study the Brownian processes (2.2) and take all coefficients like trends, dispersions and rates as constant. Due to (2.4) let's replace assumptions (2.1, 2.2) on the price $p(t)$ by assumptions on the Brownian motion $dW_c$ of the value $C(t)$ and the Brownian motion $dW_v$ of the volume $V(t)$ similar to (2.1):

$$dC(t) = C(t)[\mu_c\, dt + \sigma_c\, dW_c(t)] \tag{2.5}$$

$$dV(t) = V(t)[\mu_v\, dt + \sigma_v\, dW_v(t)] \tag{2.6}$$

$$<dW_c(t)> = 0; \quad <dW_c(t)dW_c(t)> = dt \tag{2.7}$$

$$<dW_v(t)> = 0; \quad <dW_v(t)dW_v(t)> = dt \tag{2.8}$$

Let's take that Brownian processes $dW_c$ and $dW_v$ are correlated as:

$$<dW_c(t)dW_v(t)> = \lambda\, dt \tag{2.9}$$

Taking into consideration Ito's lemma (2.5, 2.7, 2.10) define $dp(t)$ of the price $p(t)$ (2.4) as:

$$dp(t) = d\frac{C(t)}{V(t)} = p(t)\left[\frac{dC(t)}{C(t)} - \frac{dV(t)}{V(t)} + \frac{dV(t)}{V(t)}\frac{dV(t)}{V(t)} - \frac{dC(t)}{C(t)}\frac{dV(t)}{V(t)}\right] \quad (2.10)$$

It is easy to show that *dp(t)* takes form:

$$dp(t) = p(t)[\mu_p dt + (\sigma_c\, dW_c(t) - \sigma_v dW_v(t))] \quad (2.11)$$

$$\mu_p = \mu_c - \mu_v + \sigma_v^2 - \sigma_c\, \sigma_v \lambda$$

The form of the mean price trend $\mu_p$ causes no impact on BSM equations. The presence of two Brownian processes $dW_c$ and $dW_v$ in (2.1) generates main impact on BSM-like equation. Option price *S* should depend on value *C* (2.5) and volume *V* (2.6) random processes or on price *p* (2.11) and volume *V* (2.6). Many papers present option pricing under action of multiple Brownian processes (Broadie and Detemple 1997; Rapuch and Roncalli 2004; Carmona and Durrleman 2006; Li, Deng and Zhou 2010; Hull and White 1987; Heston 1993; Ball and Roma 1994; Britten-Jones and Neuberger 2000; Cohen and Tegner 2018). We don't aim the derivation of option pricing equation under action of two Brownian processes. For the simplest assumptions of classical BSM model we demonstrate that randomness of the value *C* and the volume *V* of market transactions (2.4) induces 2-dimensional option pricing equation that depend on price *p* (2.11) and volume V (2.6). It is easy to show (Hull 2009; Poon 2005) that (2.6-2.9; 2.11) cause:

$$\frac{\partial S}{\partial t} + rp\frac{\partial S}{\partial p} + rV\frac{\partial S}{\partial V} + \frac{1}{2}p^2\sigma^2\frac{\partial^2 S}{\partial p^2} + \frac{1}{2}V^2\sigma_v^2\frac{\partial^2 S}{\partial V^2} + pV\varrho\frac{\partial^2 S}{\partial V \partial p} = rS \quad (2.12)$$

$$\sigma^2 = \sigma_c^2 - 2\lambda\sigma_c\sigma_v + \sigma_v^2 \ ; \ \varrho = \lambda\sigma_c\sigma_v - \sigma_v^2 \quad (2.13)$$

We use risk-free portfolio *Π* and risk-free rate *r*.

$$\Pi(t) = S - \alpha p - \beta V \ ; \quad d\Pi = r\Pi dt \quad (2.14)$$

The derivation of (2.12; 2.13) is standard (Hull, 2009; Poon, 2005) and we omit it here. Equation (2.12) describes option price dynamics on 2-dimensional space *(p,V)*. It is easy to show that if value *C* (2.5) and volume *V* (2.6) follow identical Brownian motion *dW* and (2.9) *λ=1* then equation (2.12) takes form of 1-dimensional BSM equation (2.3).

The same reduction from (2.12) to (2.3) follows if the volume *V(t)* is a regular function or constant. Thus classical BSM model describes option pricing in the assumption that the value *C* and the volume *V* of the market transactions follow the identical Brownian motion or the value *C* or volume *V* are regular functions or constant. We omit here the change of variables that leads to the equation as it is simple and gives no new meaning.

If the value *C(t)* (2.5) and the volume *V(t)* (2.6) are described by different Brownian processes and the price *p(t)* follows (2.11) then the option price *S* should obey the 2-dimensional BSM-like equation (2.12).

## 3. Expectations and option pricing

As we show above the simple relations (2.4) enlarge the "space" of the BSM equation from one to two dimensions. However economic agents perform market transactions and agents take decisions on the value, the volume and the price of the transactions under personal expectations. Agents expectations approve market transactions and impact evolution of the value, the volume and the price of transactions. Different agents may take their decisions on base of different expectations and random perturbations of numerous expectations may cause random disturbances of the value, the volume and the price of transactions. Studies of expectations and their impact on economic and financial markets have a long history and we mention only some starting with Keynes (1936), Muth (1961) and Lucas (1972) and further research by (Sargent and Wallace 1976; Hansen and Sargent 1979; Kydland and Prescott 1980; Blume and Easley 1984; Greenwood and Shleifer 2014; Manski 2017). Usually expectations are treated as agents forecasts of trends and values of economic and financial variables, inflation and bank rates, income and prices, technology and weather forecasts and etc. We regard agents expectation as their assumptions on future state and dynamics of any economic variables or factors that can impact economic development. Variability and diversity of factors and variables that establish agents expectations make them the major source of the randomness that impact decisions on market transactions and through them the major stochastic impact on price dynamics. Expectations are most ambiguous economic issues. To simplify the problem as much as possible let's take the following assumptions. Let's propose that each market transaction is performed under expectations those approve decisions on the value and the volume of transactions taken by two agents involved into transaction. Let's define as $x_j$, $j=1,..4$ expectations of agents involved into market transaction. Let's take that $x_1$, $x_2$ – describe expectations on the value and the volume of the first agent – the seller and $x_3$, $x_4$ describe expectations on the value and the volume of the second agent – the buyer. Let's assume that expectations impact each other and the value and the volume of the transaction depend on all expectations of both agents. For simplicity let's take that random expectations $dx_j$ follow the Brownian motion

$$dx_j = \mu_j dt + \sigma_j dW_j \quad ; \quad j = 1,..4 \quad (3.1)$$

$$<dW_i dW_j> = \lambda_{jk} dt \quad ; \lambda_{jj} = 1 ; |\lambda_{jk}| \leq 1 ; \quad j,k = 1,..4 \quad (3.2)$$

We study the case when random expectations $dx_j$ cause change of the value $C$ and the volume $V$ of the transactions as:

$$dC = C \sum_{j=1,.4} A_j \, dx_j \quad ; \quad dV = V \sum_{j=1,.4} B_j \, dx_j \qquad (3.3)$$

Relations (3.3) model the dependence of the value $C$ and the volume $V$ of transactions on random expectations $x_j$ (3.1, 3.2). As we discussed above agent's expectations may be based on any economic or financial variables or factors those impact economic development and may determine decisions on market transactions with underlying assets. Taking into account (3.1-3.3), Ito's lemma and relations (2.10) obtain:

$$dp(t) = p\left[\mu_p dt + \sum_{j=1,.4} D_j \, dW_j\right] \quad ; \quad D_j = (A_j - B_j)\sigma_j \qquad (3.4)$$

The mean price trend $\mu_p$ is constant and its form is determined by (2.10; 3.1-3.3). The exact form of mean price trend $\mu_p$ doesn't impact BSM-like equation and we omit it for simplicity. Four random expectations $x_j$, $j=1,2,3,4$ (3.1, 3.2) imply that underline price $p(t)$ (3.4) and option price $S$ should depend on four variables. Let's take price $p$ (3.4) and three expectations $x_j$, $j=1,2,3$ (3.4; 3.5) as independent variables. Taking Ito's lemma for the case of 4 independent variables obtain the BSM-like equation on option price $S=S(t,p,x_1,x_2,x_3)$:

$$\frac{\partial S}{\partial t} + rp\frac{\partial S}{\partial p} + rx_j\frac{\partial S}{\partial x_j} + \frac{1}{2}p^2\sigma_p^2\frac{\partial^2 S}{\partial p^2} + \frac{1}{2}\sum_{j,k=1,2,3}\lambda_{jk}\sigma_j\sigma_k\frac{\partial^2 S}{\partial x_j \partial x_k} + p\varrho_j\frac{\partial^2 S}{\partial x_j \partial p} = rS \qquad (3.5)$$

$$\sigma_p^2 = \sum_{j,k=1,.4}\lambda_{jk}\, D_j D_k \qquad (3.6)$$

$$\varrho_j = \sum_{k=1,.4}\sigma_j\lambda_{jk}D_k \quad ; \quad j=1,2,3. \qquad (3.7)$$

It is clear that the seller and the buyer may perform market transactions with underlying asset on base of numerous expectations. Agents expectations are most uncertain and most influential economic issues that deliver major randomness to financial markets and economics as a whole. Ensemble of these expectations and their disturbances deliver additional uncertainty to asset price $p(t)$ and through it to option pricing. In (Olkhov 2019) we present a simple model that describes direct impact of small fluctuations of expectations on price and return fluctuations and price-volume relations. Similar model can be used to model impact of ensemble of expectations on option pricing.

## 4. Non-linear option pricing models

Expectations that govern the underlying asset transactions may concern options pricing dynamics. In other words – market traders may perform transactions with underlying assets on base of information and assessments of corresponding option trading data. Actually we believe that professional investors and traders use all market information available to them to establish their expectations and to take the preferable market transaction. Thus the value $C$ and the volume *V* of market transactions with underlying may depend on agents expectations and coefficients (3.3) may depend on agents expectations formed by current option price *S* or its derivatives by time *t* or by price *p* and etc. For example, relations (3.3) that describe direct dependence on option price *S* may model nonlinear dependence of option pricing models. Non-linear option pricing models are studied for more than 25 years (Bensaid, et.al. 1992; Sircar and Papanicolaou 1998; Frey 2008; Frey and Polte 2011; Loeper 2018).

Expectations of investors and traders on option pricing data impact their market transactions on underlying assets and cause non-linear option pricing equations. For simplicity as a toy model let's study dependence of the value *C(t)* of transactions on single expectation *x* determined by the option price *S* and *dV* in (3.3) is regular function or *dV=0*. Let's assume that coefficient *A* in (3.3) depends on expectation of the option price *S* and *A=A(S)*. Then (3.3) takes form (4.1):

$$dC = C\ A(S)dx \qquad (4.1)$$

Let's assume that investors and traders forecast change *dx* of their expectation of option price *S* due to Brownian motion (3.1) or (4.2):

$$dx = \mu dt + \sigma dW\ ;\ <dWdW> = dt \qquad (4.2)$$

Then, for single expectation *x* (3.1; 3.3; 4.2) and due to (2.10) the price *dp* takes form:

$$dp(t) = \frac{dC}{V} = p(t)A(S)dx = p(t)A(S)[\mu dt + \sigma dW] \qquad (4.3)$$

and the equation (2.12) is transformed to classical BSM equation (2.3) with non-linear term:

$$\frac{\partial S}{\partial t} + rp\frac{\partial S}{\partial p} + \frac{1}{2}A^2(S)\sigma^2 p^2 \frac{\partial^2 S}{\partial p^2} = rS \qquad (4.4)$$

We don't study here any particular non-linear BSM-like equation but outline a simple and direct way to take into the consideration the impact of the expectations on the option pricing and method for derivation of corresponding the non-linear BSM-like equations. The similar assumptions can induce more sophisticated non-linear BSM-like equations in two, three or four dimensions that take into account impact of two, three or four expectations starting with (3.5). Such non-linear BSM-like equations can describe the dependence of coefficients $A_j$, $B_j$ (3.3) on the option price *S* or its derivatives by time *t* or by price *p*.

Description of the impact of expectations on transactions and their value *C* and volume *V* is rather complex problem. There are numerous agents involved into market transactions with underlying or with options. Different agents perform their transactions under numerous expectations. As we mentioned above agents may establish their expectations on base of any economic and financial variables, market and tax trends, technology and climate forecasts, and on base of any social or psychology factors that may impact agent's mood. Thus description of option pricing as well as description of asset pricing should take into account definite "mean" action of various expectations that impact decisions of different agents. Such "mean" expectations as well as fluctuations from "mean" expectations impact underlying and option pricing. The methods for description of distribution of expectations of different agents and modeling "mean" expectations are presented in Olkhov (2019). These methods introduce distributions of agents, transactions and expectations that help describe price-volume and return-volume disturbances for asset pricing. The approach to economic modeling developed by (Olkhov 2016a; 2016b) gives opportunity to discuss some hidden problems of option pricing (Olkhov 2016c). We refer for these studies for further details.

## 5. Stochastic volatility

It is well known that assumption on constant volatility of the classical BSM model doesn't match the market reality. Numerous extensions of BSM equations were proposed to describe impact of stochastic volatility on option pricing. Stochastic volatility models were developed starting at least with Cox and Ross (1976) and then followed by Hull and White (1987), Heston (1993), Ball and Roma (1994), Saikat (1996), Poon (2005), Engle and Figlewski (2014), Cohen and Tegner (2018). Further studies of stochastic volatility models concern usage of various assumptions on properties of stochastic processes that may describe real properties of market volatility variations. Stochastic volatility models (Heston 1993, Poon 2005) describe transition from 1-dimensional BSM equation to 2-dimensional heat-type equation. We show that impact of random the value and the volume of transactions with underlying induces 2-dimensional BSM-like equation (2.12; 2.13). If one takes into account impact of expectations those approve decisions on market transactions then option pricing may obey two, three or four-dimensional BSM-like equations with constant volatilities. Extensions of equations (2.12; 2.13) to stochastic volatility model introduce two additional random variables: random value volatility $\sigma^2_c$ and random volume volatility $\sigma^2_v$ that follow Brownian motion $dW_{\sigma c}$ and $dW_{\sigma v}$. Let's define

$$x = \sigma_c^2 \quad ; \quad y = \sigma_v^2 \qquad (5.1)$$

Relations (2.5-2.9) those define equation (2.12) stochastic volatility model are complemented by additional relations (Heston, 1993; Poon, 2005)

$$dx = \alpha_x(\theta_x - x)dt + \sigma_x\sqrt{x}\,dW_x \qquad (5.2)$$

$$dy = \alpha_y(\theta_y - y)dt + \sigma_y\sqrt{y}\,dW_y \qquad (5.3)$$

Relations (2.5-2.9) and (5.1; 5.2) define four independent Brownian motions and induce corresponding 4-dimension BSM-like equation. To avoid excess complexity let's present 3-dimensional Heston-like equation $S=S(t,p,V,x)$ that model the stochastic value volatility $x=\sigma^2_c$ and keep the volume volatility $y=\sigma_v^2$ - const.

$$\frac{\partial S}{\partial t} + rp\frac{\partial S}{\partial p} + rV\frac{\partial S}{\partial V} + [\alpha_x(\theta_x - x) - \vartheta x]\frac{\partial S}{\partial x} + \frac{1}{2}p^2(x - 2\lambda\sigma_v\sqrt{x} + \sigma_v^2)\frac{\partial^2 S}{\partial p^2} + \frac{1}{2}V^2\sigma_v^2\frac{\partial^2 S}{\partial V^2} +$$
$$\frac{1}{2}\sigma_x^2 x\frac{\partial^2 S}{\partial x^2} + p\sigma_x\sqrt{x}\left[\sqrt{x}\lambda_{cx} - \sigma_v\lambda_{vx}\right]\frac{\partial^2 S}{\partial x\partial p} + V\sigma_v\sigma_x\sqrt{x}\,\lambda_{vx}\frac{\partial^2 S}{\partial V\partial x} + pV\varrho\frac{\partial^2 S}{\partial V\partial p} = rS \quad (5.4)$$

$$<dW_x(t)dW_c(t)> = \lambda_{xc}\,dt \quad;\quad <dW_x(t)dW_v(t)> = \lambda_{xv}\,dt \quad (5.5)$$

If $dW_v$ is identical to $dW_c$ or if $V$-const then the equation (2.12) is reduced to the classical (2.3) and (5.4) is reduced to the Heston stochastic volatility equation (5.6) (Heston 1993)

$$\frac{\partial S}{\partial t} + rp\frac{\partial S}{\partial p} + [\alpha_x(\theta_x - x) - \vartheta x]\frac{\partial S}{\partial x} + \frac{1}{2}p^2 x\frac{\partial^2 S}{\partial p^2} + \frac{1}{2}\sigma_x^2 x\frac{\partial^2 S}{\partial x^2} + \lambda_{xc}\sigma_x px\frac{\partial^2 S}{\partial x\partial p} = rS \quad (5.6)$$

During more then 25 years transitions of the option pricing from constant to stochastic volatility models in the Heston approximation were described in numerous papers (Heston 1993; Poon 2005; Cohen and Tegner 2018) and we are not going to reproduce them once more. We just show that description of stochastic volatility of the value and the volume of transactions or stochastic volatility of expectations increase dimension of the equations (2.12) and (3.6) and add extra complexity for option pricing modeling. Nonlinear equations (4.4) with constant volatilities $\sigma^2$ also can be starting points for extension to stochastic volatility approximation. Description of stochastic volatility increase dimension of the equation (4.4) and add extra complexity for solving high dimensional nonlinear equations.

## 6. Discussion

Let's discuss some internal problems of option pricing modeling. Studies of these problems may clarify relations between forecasting of option pricing on base of the BSM-like equations and real market data.

High frequency trading can deliver thousands records of market transactions per second. This information is very useful for short-term market assessment during one hour or one day. But even intraday volatility assessments may require initial aggregation of high frequency market data for seconds, minutes or hours. Option pricing assessments for weeks and months may need time data aggregation during hours or days. Aggregation of market data by time term $t_1$ means that the option price evolution model has internal time scale $t_1$. This time scale $t_1$ may impact random properties of the underlying price and impact option price model dynamics. The second time scale $t_2 > t_1$ is responsible for the averaging procedure $<..>$ to assess the mean value and volatility of Brownian processes (1.2). The scale $1/t_2$ define maximum frequency scale for the problem under consideration. Usage of different time scales $t_1$, $t_2$ to solve the same options pricing problem with scale - time to maturity $T$ may cause different properties, as distinctive frequency scales will be different. On the other hand usage of time scales $t_1$ and $t_2$ causes aggregation of economic and financial variables of the problem. In particular it means aggregation of the value and the volume of the transactions with underlying during time term $t_2$ to obtain correct values for price $p(t)$ due to relations (2.10; 2.12). Roughly speaking, one should measure the sum of the value $C(t)$ and sum of the volume $V(t)$ of all transactions during time term $t_2$ and their ratio (2.4) should define the mean price $p(t)$. Hence the solution $S(T,p,V)$ of the option price equation (2.12) should depend on the internal time scale $t_2$.

Existence of the internal time scales $t_1 < t_2$ for the option pricing and the requirement to use relations (2.4, 2.10) to define the price $p(t)$ at moment $t$ arises the problem of impact of expectations on asset and option pricing. Indeed, relations (2.4) for the value $C(t)$ and the volume $V(t)$ aggregated by all transaction during time term $t_2$ define so-called volume weighted average price $p(t)$ (VWAP) of transactions (Berkowitz et.al 1988; Buryak and Guo, 2014; Guéant and Royer, 2014; Busseti and Boyd, 2015; Padungsaksawasdi and Daigler, 2018). However expectations of agents those approve agents decisions on the value of the volume of the performed transactions can be based on any factors that might impact agents will to make a deal. For example, agents expectations can be based on assessment of price $\pi(t)$ as simple average during same time term $t_2$. The difference between VWAP and assessments of price $\pi(t)$ as simple average may be great. Thus agents expectations may reflect price assessments based on different models and methods and that induce additional disturbances and stochastic to the price results (1.1) of the market transactions.

Above issues and variety forms and dimensions of possible linear or nonlinear equations (2.12; 3.6; 4.4; 5.4) that can model options pricing in different assumptions on constant or stochastic volatility or on impact of expectations arise the problem of sufficiency requirements of option modeling. The classical BSM equations (2.3) deliver simple 1-dimensional model that describe the option price dynamics with reasonable accuracy. Classical BSM model has only two parameters. One should only wonder how such a simple equation (2.3) gives so good model forecast for option pricing. Any further extensions of the classical BSM model add more accuracy but that cost extra complexity.

We consider the classical BSM model assumptions and replace the only one. We take into account the dependence of underlying price on the value and the volume of the market transactions. This trivial point gives sufficient reasons to extend the 1-dimensional BSM equations (2.3) to 2-dimensional equations (2.12). But such extension even for the case with constant parameters costs extra volume volatility and correlation coefficients. Transition from constant to stochastic volatility approximations turn the model to much more complex 3- or 4-dimensional equations (5.4).

Common understanding that market transactions are governed by agents expectations and infinite diversity of these expectations transfers the relatively simple 2-dimensional equation (2.12) to 4-dimensional equation (3.6) with constant coefficients. Any attempt to model stochastic volatility of underlying asset price of stochastic volatility of expectations would increase the dimension and the complexity of the equations. Moreover any sophistication of option pricing equations requires additional sophistication of market econometrics. Actually market data on the value and the volume of executed transactions are among the usual and we hope that we don't add excess complexity to option price description on base of (2.12). As well the attempt to take into considerations the impact of agents random expectations on market transactions and use corresponding econometric data may be not too simple.

We regard the balance between the accuracy of the description and complexity of econometrics as main problem for asset and option pricing modeling. We don't see any place for unique "correct and perfect" option pricing model equation but propose that different approximations should serve for different option markets. Any step towards accuracy will be compensated by two steps back to complexity due to changes in agents expectations and their impact on asset and option pricing. Nonlinear relations even in simplest form modeled by (4.4) reflect the tip of the complex mutual dependence between underlying and options.

## 7. Conclusion

After almost 50 years since publication the classical BSM options pricing model remains the source for further studies. We reconsider the BSM model and consistently use the price definition (1.1) as the only correct price time-series records determined by executed market transactions. The relations (1.1) present the price $p$ as coefficient between the value $C$ and the volume $V$ of the market transaction. The usage of simple relations (1.1) for the BSM option pricing model extends the BSM equation (1.3) from one to two dimensional BSM-like equation (4.1). As we show in Sec.2 the classical BSM one-dimensional model implicitly use the assumption that the value $C$ and the volume $V$ follow the identical Brownian motion. The deviation of the value $C$ and the volume $V$ from identical Brownian processes leads to 2-dimensional BSM-like model (2.12). Any price definitions, models and forecasts different from (1.1) may have sense as ground for agents price expectations only. It is clear that agents price expectations those approve execution of the market transactions impact the price dynamics (1.1). Nevertheless relations (1.1) remain the only correct set of price records of executed market transactions. The influence of agents price expectations on option pricing lead to further complicating and increase of dimension of the option pricing models. We present simple models that take into account impact of agents expectations on performance of market transactions. Agents may go into transactions under their individual expectations on the value, volume or price dynamics or any other expectations. As we show in Sec. 3 and 4 taking into account agents expectations can further increase the dimension of the BSM model or give further complication for non-linear option model.

Equation (3.6) serves as starting point for modeling (4.4) non-linear relations between underlying and option market and opens the way for equation (5.4) that describe the Heston stochastic volatility model impact on option pricing. The set of the BSM-like equations (2.12; 3.6; 4.4; 5.4) have only common origin: the simple relations (1.1) that determine the price $p$ as the coefficient between the value $C$ and the volume $V$ of the market transactions and assumption that agents go into transactions under their individual expectations.

These equations don't simplify the description of the option pricing but may adopt impact of the real economic and financial factors on the underlying and the option pricing. The most difficult problem of all further extensions is the achieving the balance between the next level of accuracy of the asset and option pricing model and the next level of complexity. Complex market relations require complex equation to describe real market processes and we hope that usage of equations (2.12; 3.6; 4.4; 5.4) and obvious directions for their extensions may improve current asset and option pricing modeling.

## Acknowledgements

We thank anonymous referees for their reasonable comments on Ito's lemma usage and language corrections.